\documentclass[12pt,preprint,longabstract]{aastex}
\usepackage{color}
\usepackage{amsmath}
\shortauthors{KamLAND collaboration}

\begin{document}
\title{Study of electron anti-neutrinos associated with
 gamma-ray bursts using KamLAND}

\author{
K.~Asakura\altaffilmark{1}, 
A.~Gando\altaffilmark{1}, 
Y.~Gando\altaffilmark{1}, 
T.~Hachiya\altaffilmark{1},  
S.~Hayashida\altaffilmark{1}, 
H.~Ikeda\altaffilmark{1},
K.~Inoue\altaffilmark{1,2},
K.~Ishidoshiro\altaffilmark{1},
T.~Ishikawa\altaffilmark{1}
S.~Ishio\altaffilmark{1}
M.~Koga\altaffilmark{1,2},
S.~Matsuda\altaffilmark{1},
T.~Mitsui\altaffilmark{1},
D.~Motoki\altaffilmark{1},
K.~Nakamura\altaffilmark{1,2}, 
S.~Obara\altaffilmark{1}, 
Y.~Oki\altaffilmark{1},  
T.~Oura\altaffilmark{1}, 
I.~Shimizu\altaffilmark{1}, 
Y.~Shirahata\altaffilmark{1}, 
J.~Shirai\altaffilmark{1},
A.~Suzuki\altaffilmark{1}, 
H.~Tachibana\altaffilmark{1},
K.~Tamae\altaffilmark{1},
K.~Ueshima\altaffilmark{1},
H.~Watanabe\altaffilmark{1},
B.D.~Xu\altaffilmark{1},
H.~Yoshida\altaffilmark{1,14},
A.~Kozlov\altaffilmark{2},
Y.~Takemoto\altaffilmark{2},
S.~Yoshida\altaffilmark{3},
K.~Fushimi\altaffilmark{4},
A.~Piepke\altaffilmark{5,2},
T.~I.~Banks\altaffilmark{6},
B.~E.~Berger\altaffilmark{6,2},
B.K.~Fujikawa\altaffilmark{6,2},
T.~O'Donnell\altaffilmark{6},
J.G.~Learned\altaffilmark{7},
J.~Maricic\altaffilmark{7}, 
M.~Sakai\altaffilmark{7},
L.~A.~Winslow\altaffilmark{8},
Y.~Efremenko\altaffilmark{9,10,2},
H.~J.~Karwowski\altaffilmark{11},
D.~M.~Markoff\altaffilmark{11},
W.~Tornow\altaffilmark{11,2},
J.~A.~Detwiler\altaffilmark{12,2},
S.~Enomoto\altaffilmark{12,2},
M.P.~Decowski\altaffilmark{13,2}
}

\affil{The KamLAND Collaboration}

\altaffiltext{1}{Research Center for Neutrino Science, Tohoku University, Sendai 980-8578, Japan}
\altaffiltext{2}{Kavli Institute for the Physics and Mathematics of the Universe (WPI), University of Tokyo, Kashiwa 277-8568, Japan}
\altaffiltext{3}{Graduate School of Science, Osaka University, Toyonaka, Osaka 560-0043, Japan}
\altaffiltext{4}{Faculty of Integrated Arts and Science, University of Tokushima, Tokushima, 770-8502, Japan}
\altaffiltext{5}{Department of Physics and Astronomy, University of Alabama, Tuscaloosa, Alabama 35487, USA}
\altaffiltext{6}{Physics Department, University of California, Berkeley, California 94720, USA 
and Lawrence Berkeley National Laboratory, Berkeley, California 94720, USA}
\altaffiltext{7}{Department of Physics and Astronomy, University of Hawaii at Manoa, Honolulu, Hawaii 96822, USA}
\altaffiltext{8}{Massachusetts Institute of Technology, Cambridge, Massachusetts 02139, USA}
\altaffiltext{9}{Department of Physics and Astronomy, University of Tennessee, Knoxville, Tennessee 37996, USA}
\altaffiltext{10}{National Research Nuclear University, Moscow, Russia}
\altaffiltext{11}{Triangle Universities Nuclear Laboratory, Durham, North Carolina 27708, USA; Physics Departments at Duke
University, Durham, North Carolina 27705, USA; North Carolina Central University, Durham, North Carolina
27701, USA and the University of North Carolina at Chapel Hill, Chapel Hill, North Carolina 27599, USA}
\altaffiltext{12}{Center for Experimental Nuclear Physics and Astrophysics, University of Washington, Seattle, Washington 98195, USA}
\altaffiltext{13}{Nikhef and the University of Amsterdam, Science Park, Amsterdam, the Netherlands}
\altaffiltext{14}{Current address: Graduate School of Science, Osaka University, Toyonaka, Osaka 560-0043, Japan}

\begin{abstract}
We search for electron anti-neutrinos ($\overline{\nu}_e$) from long and short-duration 
gamma-ray bursts~(GRBs) using data taken by the KamLAND detector from August 2002 to June 2013. 
No statistically significant excess over the background level is found. 
We place the tightest upper limits on $\overline{\nu}_e$ fluence from GRBs 
below 7\,MeV and place first constraints on the 
relation between $\overline{\nu}_e$ luminosity and effective temperature. 
\end{abstract}
\keywords{Neutrino, GRB, KamLAND}

\section{Introduction}
Gamma-ray bursts (GRBs) are the most luminous phenomena in the universe. 
The duration of GRBs~($\Delta t_{\rm GRB}$) varies in the range between 10~ms and 1000~s, 
with a roughly bimodal distribution for so-called long GRBs of $\Delta t_{\rm GRB} \ga$ 2~s 
and short GRBs of $\Delta t_{\rm GRB} \la 2$~s. 
The progenitors of most short GRBs are widely thought to be mergers of neutron star-neutron star or black hole-neutron star binaries~\citep{meszaros2006}.  
A favored model of long GRB progenitors is a catastrophic collapse of a massive star into a black hole~\citep{meszaros2006}.  
These models are supported by observations of afterglows and identification of host galaxies for short GRBs~\citep{villasenor2005,fox2005}, 
and observations of supernovae associated with long GRBs~\citep{woosley1999,hjorth2003}. 
Both scenarios would result in the formation of a compact rotating black hole 
with an accretion disk at MeV or higher temperatures, which generates collimated relativistic fireball jets leading to GRBs. 
Although such a fireball model is promising and attractive, 
the initial condition and generation mechanism of the fireball jets are still unknown, 
since it is difficult to observe the optically thick center region of GRBs by electromagnetic waves. 

A potential scheme to directly explore the GRB center region is 
the use of thermal neutrinos and gravitational waves~(GWs)~\citep{suwa2009,sekiguchi2011_1}, 
since they have strong transmissivity. 
Thermal neutrinos are sensitive to thermodynamic profiles of the accretion disk, 
and GWs are sensitive to the dynamics of progenitors. 
Both are complementary observations to probe GRBs. 
Super-Kamiokande~(SK) and Sudbury Neutrino Observatory~(SNO) searched for MeV neutrinos related to GRBs and 
placed constraints on the upper fluence limits~\citep{sk_2,sno}. 
Others placed limits on high energy neutrinos produced in the fireball jets~\citep{amanda,sk_1,icecube,anita2011,baikal2011,Antares2013}. 
GWs from GRBs were studied by a GW detector network~\citep{ligo_network}. 

In this paper, we present a study of electron anti-neutrinos ($\overline{\nu}_e$) of a few tens of MeV energy 
produced by thermal processes from the GRB center region, especially the accretion disk~\citep{nagataki2002,sekiguchi2011,Caballero2009} 
with the Kamioka Liquid Scintillator Anti-Neutrino Detector~(KamLAND). 
We constrain the relation between the $\overline{\nu}_e$'s luminosity~($L$) and effective temperature~($T$) as well as $\overline{\nu}_e$ fluence for the first time. 
The $L$-$T$ relationship can be used to directly compare with theoretical predictions. 
These limits and constraints are established using redshift-measured GRBs. 
We adopt the standard $\Lambda$CDM cosmology 
with $\Omega_{\rm m}=0.315$, $\Omega_{\Lambda}=0.685$, and $H_0=67.3$\,km\,s$^{-1}$\,Mpc$^{-1}$~\citep{plank} throughout this paper.

\section{KamLAND detector}\label{sec2}
\begin{figure}
\begin{center}
\includegraphics[width=0.8\linewidth]{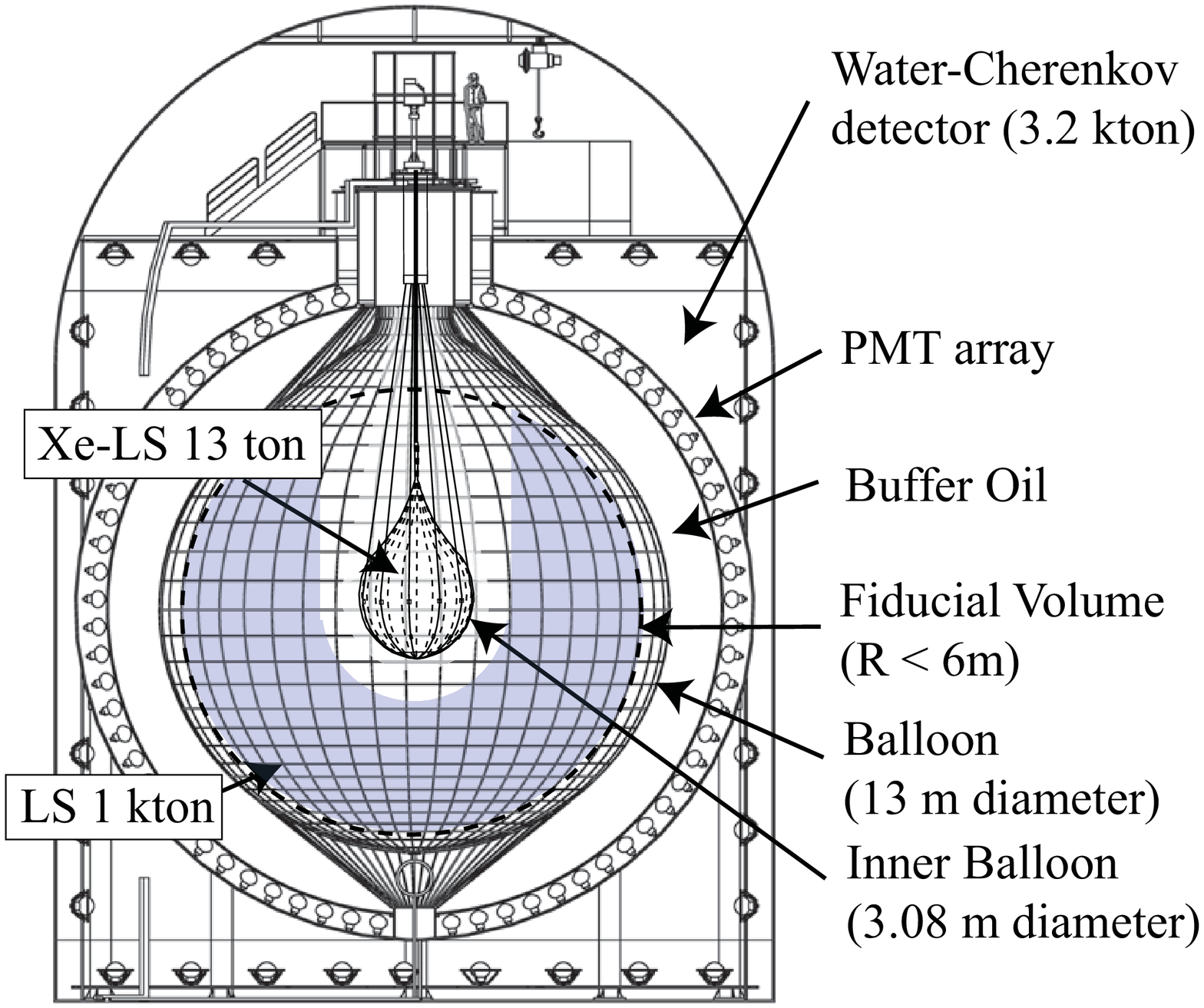}
\end{center}
\caption{Schematic diagram of the KamLAND detector. The shaded 
region in the liquid scintillator indicates the volume for 
the $\overline{\nu}_e$ analysis after the IB installation.}
\label{fig1}
\end{figure}

The KamLAND detector is located $\sim$1\,km under the peak of Mt.~Ikenoyama
($36.42^{\circ}$N, $137.31^{\circ}$E) near Kamioka, Japan.
The 2,700 meter-water-equivalent (mwe) of vertical rock overburden reduces the cosmic-ray muon flux by almost five orders of magnitude.
A schematic diagram of KamLAND is shown in Figure~\ref{fig1}.
The primary target volume consists of 1\,kton of ultra-pure liquid scintillator~(LS) contained in a 13-m-diameter spherical balloon made of
135-$\mu$m-thick transparent nylon ethylene vinyl alcohol copolymer~(EVOH) composite film.
The LS consists of 80\% dodecane and 20\% pseudocumene (1,2,4-trimethylbenzene) by volume,
and $1.36\pm0.03$~${\rm g~l}^{-1}$ of the fluor PPO (2,5-diphenyloxazole). 
A buffer comprising 57\% isoparaffin and 43\% dodecane oils by volume, which fills the region between the balloon
and the surrounding 18-m-diameter spherical stainless-steel outer vessel, shields the LS from external radiation.
The specific gravity of the buffer oil is adjusted to be 0.04\% lower than that of the LS.
An array of photomultiplier tubes~(PMTs)---1,325 specially developed fast PMTs masked to 17-inch diameter and
554 older 20-inch diameter PMTs reused from the Kamiokande experiment~\citep{Kume1983}---are mounted on the inner surface of the outer vessel, 
providing 34\% photocathode coverage.
This inner detector is shielded by a 3.2\,kton water-Cherenkov veto detector. 

KamLAND uses the inverse beta-decay reaction to detect $\overline{\nu}_e$:  
\begin{equation}
 \overline{\nu}_{e}+p\rightarrow e^{+}+n. 
\end{equation}
This process has a delayed-coincidence~(DC) event-pair signature which offers powerful background suppression. 
The energy deposited by the positron, which generates the DC pair's
prompt event, is the sum of the $e^{+}$ kinetic energy and annihilation $\gamma$ energies, $E_{\mathrm{p}}(\equiv T_{e^+} + 2m_e)$, 
and related to the incident $\overline{\nu}_e$ energy by $E_{\overline{\nu}_e} =  
\langle E_{e^+} \rangle  + \delta + E^{\rm CM}_\nu E^{\rm CM}_e/m_{\rm p}$, 
where $E^{\rm CM}_\nu$ and $E^{\rm CM}_e$ are neutrino and electron energy in the 
center of mass frame, and $\delta = (m_{n}^2-m^2_{p}-m^2_{e})/2m_{p}$~\citep{sv2003}. 
In the low energy ($E_{\overline{\nu}_e} < 20$\,MeV) range, we can approximate the above relation by 
$E_{\overline{\nu}_e} =  E_{\rm p}  + \delta E$, where $\delta E = 0.78$\,MeV. 
We use this approximation also above 20\,MeV and comment on the associated uncertainty later.
The delayed event in the DC pair is generated by a 2.2~(4.9)\,MeV $\gamma$-ray produced 
when the neutron captures on a proton~($^{12}$C).
The mean neutron capture time is $207.5 \pm 2.8~\mu$s~\citep{Abe2010}.
The angular distribution of the positron emission is nearly isotropic. 
Unlike in a water Cherenkov detector, the scintillation light is also isotropic.
As a result, the positron signal does not provide the incoming $\overline{\nu}_e$ source direction. 
Due to the extremely low cross section of $\overline{\nu}_e$, 
the Earth does not shadow MeV-energy extraterrestrial $\overline{\nu}_e$. 
The detector therefore has isotropic sensitivity to GRBs.  

The event energy and vertex reconstruction are based on the timing and charge distributions of scintillation photons recorded by the PMTs. 
The reconstruction algorithms are calibrated with on-axis and off-axis radioactive sources deployed from a glove box installed at the top of the detector.
The radioactive sources are $^{60}$Co, $^{68}$Ge, $^{203}$Hg,
$^{65}$Zn, $^{241}$Am$^{9}$Be, $^{137}$Cs, and $^{210}$Po$^{13}$C,
providing energy and vertex calibration~\citep{Berger2009,Banks201588}. 
The overall vertex reconstruction resolution is $\sim 12$~cm$/\sqrt{E( {\rm MeV})}$
and energy resolution is $6.4\%/\sqrt{E( {\rm MeV})}$.
The energy reconstruction of positrons with $E_{\rm p} > 7.5$~MeV (i.e., $E_{\overline{\nu}_e} > 8.3$~MeV) 
is verified by using tagged $^{12}$B $\beta^{-}$-decays generated via muon spallation~\citep{Abe2010}. 

In September 2011, the KamLAND-Zen double-beta ($\beta\beta$) decay search experiment 
was launched~\citep{gando2012}. 
This experiment makes use of KamLAND's extremely low background. 
The KamLAND detector was modified to include a $\beta \beta$ source, 
13\,tons of Xe-loaded liquid scintillator (Xe-LS) contained 
in a 3.08-m-diameter inner balloon~(IB), at the center of the detector. 

\section{Event selection}\label{sec3}
\subsection{KamLAND DC events}
In this analysis, we use KamLAND data collected from August 3, 2002 to June 4, 2013. 
During the majority of this period, KamLAND was measuring $\overline{\nu}_e$
from nuclear power plants with a spectrum up to about 8\,MeV~\citep{kamland2011,gando2013} 
and geological $\overline{\nu}_e$ from the Earth's deep interior~\citep{araki2005, gando2011, gando2013}. 
Following the Fukushima reactor accident in March 2011, all Japanese reactor were subject 
to a protracted shutdown. The data-set is divided into two periods. 
Period I refers to data that was taken until the IB installation in September 2011. 
Period I\hspace{-.1em}I refers to the data taken after the IB installation, 
which mostly coincided with the low reactor $\overline{\nu}_e$ flux. 

In Period I, we search only for $\overline{\nu}_e$ events with $ E^{\rm I}_{\rm low}(=7.5$\,MeV) $\leq E_{\rm p} \leq 100$\,MeV  
which corresponds to the energy range of interest for GRBs with almost zero contamination 
from the reactor $\overline{\nu}_e$ flux. 
During Period I\hspace{-.1em}I, the reactor signal is minimal, 
allowing a reduction of the energy threshold to $E^{\rm I\hspace{-.1em}I}_{\rm low}=0.9$\,MeV.  

For the DC event pair selection, we apply the following series of selection cuts: 
the prompt energy is required to be $E^k_{\rm low} \leq E_{\rm p} \leq 100$~MeV in Period $k$, 
and the delayed energy to be $1.8$~MeV $\leq E_{\rm d} \leq 2.6$~MeV for neutron capture on protons 
or $4.4$~MeV $\leq E_{\rm d} \leq 5.6$~MeV for neutron capture on $^{12}$C, 
a fiducial volume cut of $R \leq 6$~m from the center of the balloon on both prompt and delayed events, 
a spatial correlation cut of $\Delta R \leq 1.6$~m and a time separation cut of $0.5$~$\mu$s $\leq \Delta T \leq 1.0$~ms. 
Spallation cuts were used to reduce backgrounds from long-lived isotopes, e.g., $^9{\rm Li}$~($\tau = 257$~ms and $Q = 13.6$~MeV), 
that are generated by cosmic muons passing through the scintillator. 
In Period I\hspace{-.1em}I, we have to use an additional spatial cut for delayed events to avoid backgrounds 
from the IB and its support material as shown in Figure \ref{fig1}~\citep{gando2013} and 
a second-level cut using a likelihood discriminator to reduce accidental backgrounds in the low-energy region~\citep{kamland2011}. 
The selection efficiency ($\epsilon_{\rm s}^k$) is evaluated from Monte Carlo simulation separately 
for Period I ($k$=I) and Period I\hspace{-.1em}I ($k$=I\hspace{-.1em}I) due to these additional cuts. 
Note that $\epsilon_{\rm s}^{\rm I\hspace{-.1em}I}$ depends on $E_{\rm p}$ because of the energy-dependent second-level cut.  
The number of target protons in $R \leq 6$~m is estimated to be $N_{\rm T}=(5.98 \pm 0.12) \times 10^{31}$. 

The total livetime during Period I was 6.91\,yr and 55 DC events were observed during this period. 
In Period I\hspace{-.1em}I, KamLAND found 88 DC events with 1.2\,yr livetime. 
The livetime is defined as the integrated period of time that the detector was sensitive to $\overline{\nu}_e$ and 
includes corrections for calibration periods, detector maintenance, daily run switch,  etc. 
The event rates are $9.1 \times 10^{-4}$ and $8.4 \times 10^{-3}$ events per hour 
in Period I and I\hspace{-.1em}I, respectively.  
 

\subsection{GRB events}
We use GRB events observed by one or more of {\it SWIFT}, {\it HETE-2}, {\it Ulysses}, {\it INTEGRAL}, {\it AGILE}, {\it MAXI}, and {\it FERMI}  
based on The Gamma-ray Coordinates Network\footnote{\tt http://gcn.gsfc.nasa.gov/}. 
Initial selection criteria are the requirement that the GRB be in the time period between August 3, 2002 and June 4, 2013 
and the existence of redshift and GRB-duration time measurements. 
At this stage, 256 long GRB and 21 short GRB events are left. 
Subsequently, all the KamLAND runs\footnote{
KamLAND data-taking is stopped and restarted every day to ensure smooth data taking. 
The length of a KamLAND run is typically 24 hours long.} 
that include GRB events must have passed basic quality criteria (e.g., not a calibration run and stable operation). 
This leaves 175 long GRBs and 17 short GRBs in Period I. 
Period I\hspace{-.1em}I contains 38 long GRBs and one short GRB. 
One can see our GRB list by the website\footnote{see \tt http://www.awa.tohoku.ac.jp/KamLAND/GRB/2015}. 

\section{Data analysis}
\begin{figure}
\begin{center}
\includegraphics[width=0.8\linewidth]{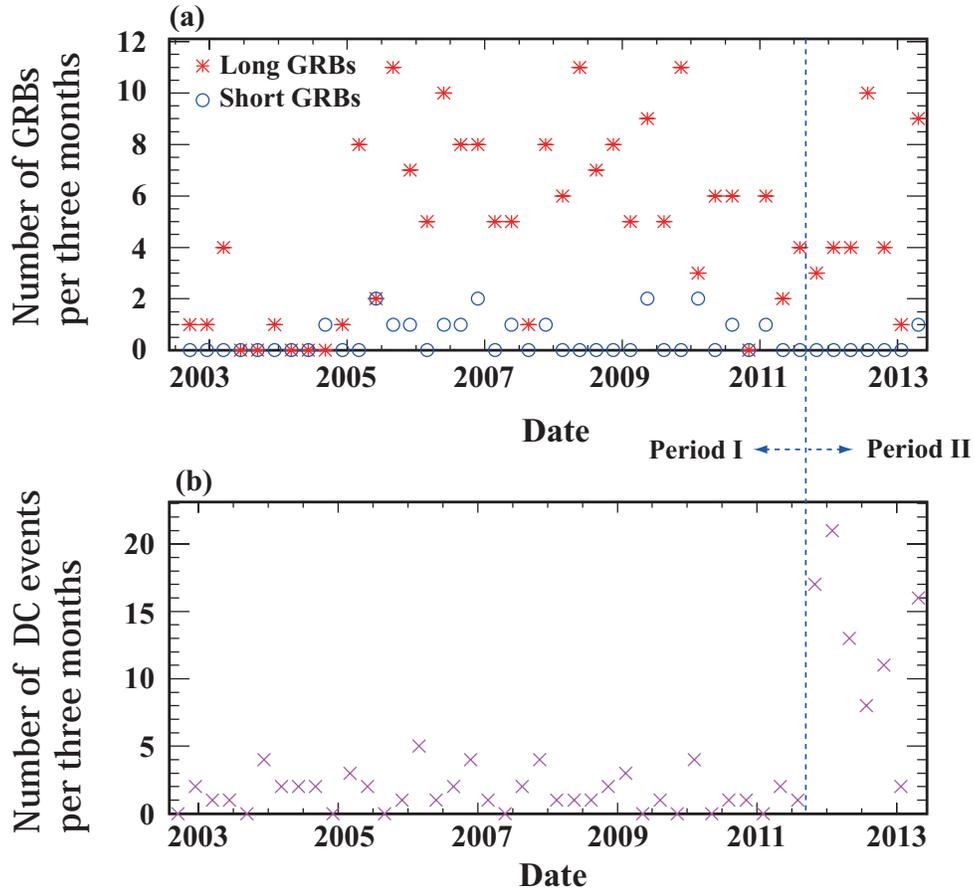}
\end{center}
\caption{Number of long and short GRBs per three months (upper panel) and 
DC events per three months (bottom panel). The number of GRBs significantly increased due to 
the SWIFT satellite after December 2004. During Period I, the DC event rate is within the statistical uncertainty.
Period I\hspace{-.1em}I started in September 2011 and allowed for a lower energy threshold, 
increasing the number of DC events.  }
\label{fig2}
\end{figure}

The average number of DC events and GRB events per three months is shown in Figure \ref{fig2}. 
In this figure, one can see a ``step'' before and after the launch of the {\it SWIFT} satellite~(November 2004) for GRB events. 
In contrast, there is no time dependence of the DC event rate during each periods. 
We therefore decided to analyze the whole KamLAND data regardless of the GRB event rate.  

\subsection{Coincidence analysis}\label{sec:coincidence_analysis}
We conduct a time-coincidence analysis between the redshift-measured GRB samples and the KamLAND DC events for long and short GRBs. 
The coincidence search time-window between a GRB event and a KamLAND DC event is defined as: 
$-t_{\rm p} + T_{\rm GRB} < T_{\rm DC} < T_{\rm GRB} + \Delta t_{\rm GRB} + t_{\rm p} + t_{\rm f}(z)$, 
where $T_{\rm DC}$ and $T_{\rm GRB}$ are the absolute times of the KamLAND DC and GRB events, respectively. 
$\Delta t_{\rm GRB}$ is the measured GRB duration time, $t_{\rm p}$ is 150\,sec corresponding to 
a model-dependent, but reasonable time difference between the thermal neutrino production 
and the GRB photon production~\citep{private,private2}, 
and $t_{\rm f}(z)$ is the relativistic flight-time delay of MeV neutrinos due to non-zero neutrino mass~\citep{li2005,choubey2003}:
\begin{equation}
t_{\rm f}(z)=\frac{1}{2}\frac{m_{\overline{\nu}_{e}}^2}{E_{\overline{\nu}_e}^2}\int^z_0 \frac{dz'}{(1+z')^2 H_0 \sqrt{\Omega_{\Lambda} + (1+z')^3\Omega_{\rm m}}},
\end{equation}
with the assumption of $m_{\overline{\nu}_e} = m_{\rm heaviest} = 87.2$~meV from $\sum m_{\nu} \leq 0.23$\,eV~\citep{plank} 
and $E_{\overline{\nu}_e} \geq 8.3$~MeV in Period I 
and $E_{\overline{\nu}_e} \geq 1.8$~MeV in Period I\hspace{-.1em}I. 
All parameters of the time-window are fixed before the coincidence search. 
The total window length for the long GRBs is 25.2 hours 
(18.3 hours in Period I and 6.82 hours in Period I\hspace{-.1em}I). 
The short GRBs sum to a total of 1.45 hours of on-time window 
(1.33 hours in Period I and 0.11 hours in Period I\hspace{-.1em}I). 

\begin{figure}
\begin{center}
\includegraphics[width=0.8\linewidth]{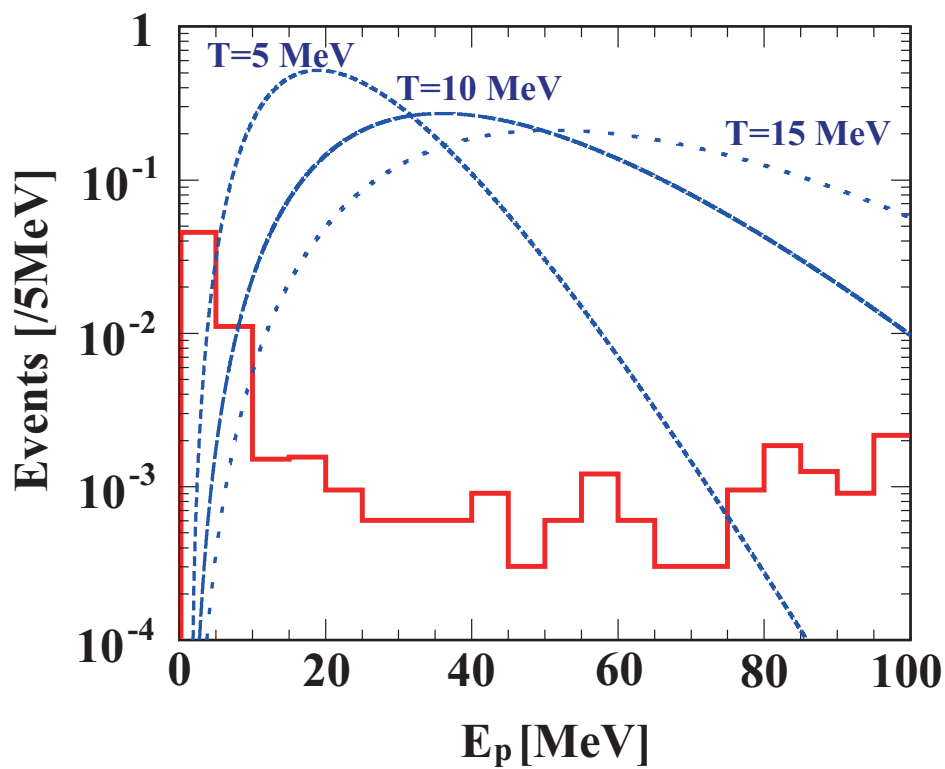}
\end{center}
\caption{
Combined background spectrum of Period I and I\hspace{-.1em}I~(solid red). 
The integral is $7.4\times 10^{-2}$ events. 
The dashed curves provide our 90\% upper limits under the assumption 
of a Fermi-Dirac distribution at temperature $T$ of $\overline{\nu}_e$. 
Each dashed curve from the left to the right corresponds to $T=5,10,15$\,MeV.} 
\label{fig3}
\end{figure}

No coincidence DC events were found in the above time window for both long and short GRBs. 
We estimate the expected accidental coincidence of DC events to be  
$7.4\times 10^{-2}$ and $2.2\times 10^{-3}$ for long and short GRBs, respectively. 
For long GBRs, the background spectrum is shown in Figure \ref{fig3} with several expected spectra 
from our 90\% upper limits (see \ref{sec:analysis}).  
In the absence of a signal, 
the Feldman-Cousins upper~(FC) limits on the DC events are $N_{90}=2.365$ and $2.435$ 
with 90\% confidence level~(CL) for long and short GRBs, respectively. 

If we use a much longer, exotic, time window, e.g., $t_{\rm p} = 6$\,h, four coincidence DC events are found for long GRBs. 
However, the expected accidental coincidence of DC events is 3.4. 
There is therefore no statistical evidence for the detection of $\overline{\nu}_e$ from long GRBs.

\subsection{Fluence upper limits}
There is no established neutrino production model for GRBs. 
We translate our FC limits to 
model-independent upper limits on $\overline{\nu}_e$ fluence, $\Psi(E_{\overline{\nu}_e})$, 
at the detector using a Green's function, 
which represents the upper limits on monoenergetic neutrinos at that specific energy. 
We use the same methodology to estimate $\Psi(E_{\overline{\nu}_e})$  
as SK~\citep{sk_2} and SNO~\citep{sno}:  
\begin{equation}
 \Psi(E_{\overline{\nu}_e}) = \frac{N_{90}}{ \sum_{k} N^{k}_{\rm GRB} I_k(E_{\overline{\nu}_e}) },
\end{equation}
where $N^{k}_{\rm GRB}$ is the number of GRBs and $I_k(E_{\overline{\nu}_e})$ is the effective number 
of DC events per one GRB with a monoenergetic spectrum in the period $k$: 
\begin{equation}\label{eq1}
 I_k(E_{\overline{\nu}_e})=N_{\rm T}\int_{E^k_{\rm low}}^{100\,{\rm MeV}} \epsilon_t \epsilon^k_s(E^{\rm vis}_{\rm p}) \sigma_{\rm IBD}( E_{\overline{\nu}_e} ) 
\delta(E^{\rm exp}_{\rm p} + \delta E - E_{\overline{\nu}_e}) 
R(E^{\rm exp}_{\rm p},E^{\rm vis}_{\rm p})dE^{\rm exp}_{\rm p}dE^{\rm vis}_{\rm p}, 
\end{equation}
and 
\begin{equation}
 R(E^{\rm exp}_{\rm p},E^{\rm vis}_{\rm p})=\frac{1}{\sqrt{2\pi} \sigma(E^{\rm exp}_{\rm p})} \exp\Biggl( - \frac{ (E^{\rm exp}_{\rm p}-E^{\rm vis}_{\rm p})^2 }{2\sigma^2(E^{\rm exp}_{\rm p})} \Biggr). 
\end{equation}
$\epsilon_t$ is the mean livetime-to-runtime ratio\footnote{Runtime is the total time of data taking. } 
and $E^{\rm exp}_{\rm p}$ and $E^{\rm vis}_{\rm p}$ are the expected and measured prompt energies, respectively.  
$\sigma_{\rm IBD}(E)$ is the differential cross section of the inverse beta decay. 
$\sigma(E)$ corresponds to the energy resolution of $6.4\%/\sqrt{E( {\rm MeV})}$.

\begin{figure}
\begin{center}
\includegraphics[width=0.8\linewidth]{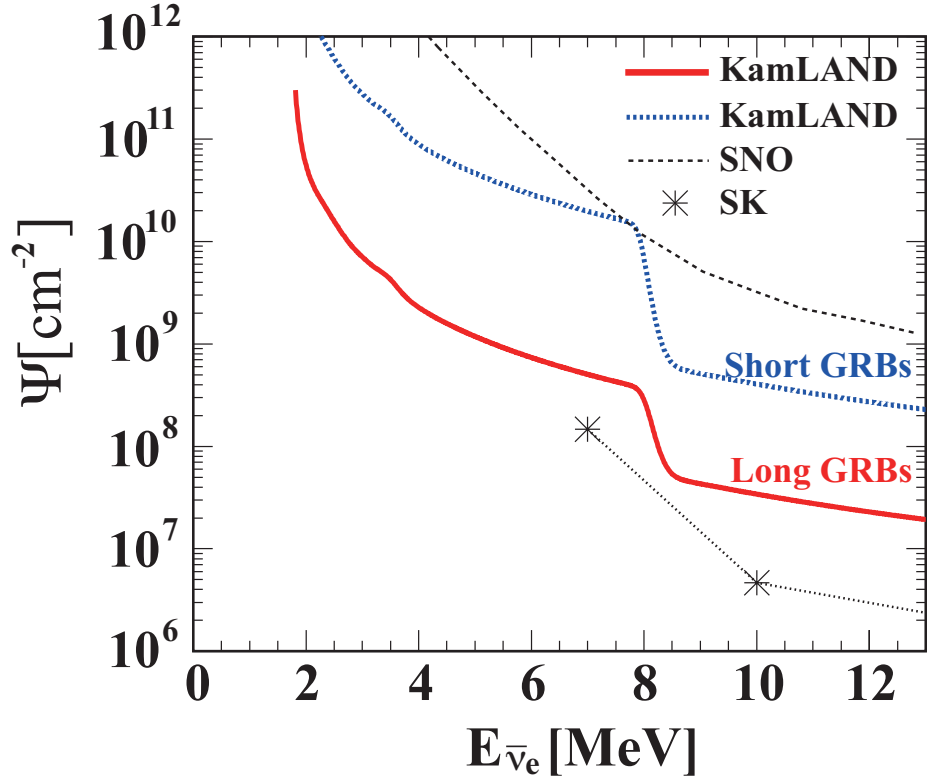}
\end{center}
\caption{Fluence upper limits on $\overline{\nu}_e$ from GRBs as a function of neutrino energy. 
Results from SK and SNO are also presented for comparison. 
SK is expected to have poorer sensitivity at low energies 
due to the detection threshold. Below 7\,MeV, KamLAND establishes the tightest limits on 
neutrino fluence. The slight distortion around 3--4\,MeV is 
from the energy dependence of the selection efficiency ($\epsilon_{\rm s}^{\rm I\hspace{-.1em}I}$). }
\label{fig4}
\end{figure}

The 90\% CL upper limits on $\Psi(E_{\overline{\nu}_e})$ from KamLAND are shown 
for both long and short GRBs together in Figure \ref{fig4} with the results from SK~\citep{sk_2} and SNO~\citep{sno}. 
Note, the results from SK and SNO treated long and short GRBs as the same. 
Below 7\,MeV, our analysis provides the best limits so far.   

\subsection{Constraint on luminosity and effective temperature ($L$-$T$)}\label{sec:analysis}
\begin{figure}
\begin{center}
\includegraphics[width=0.8\linewidth]{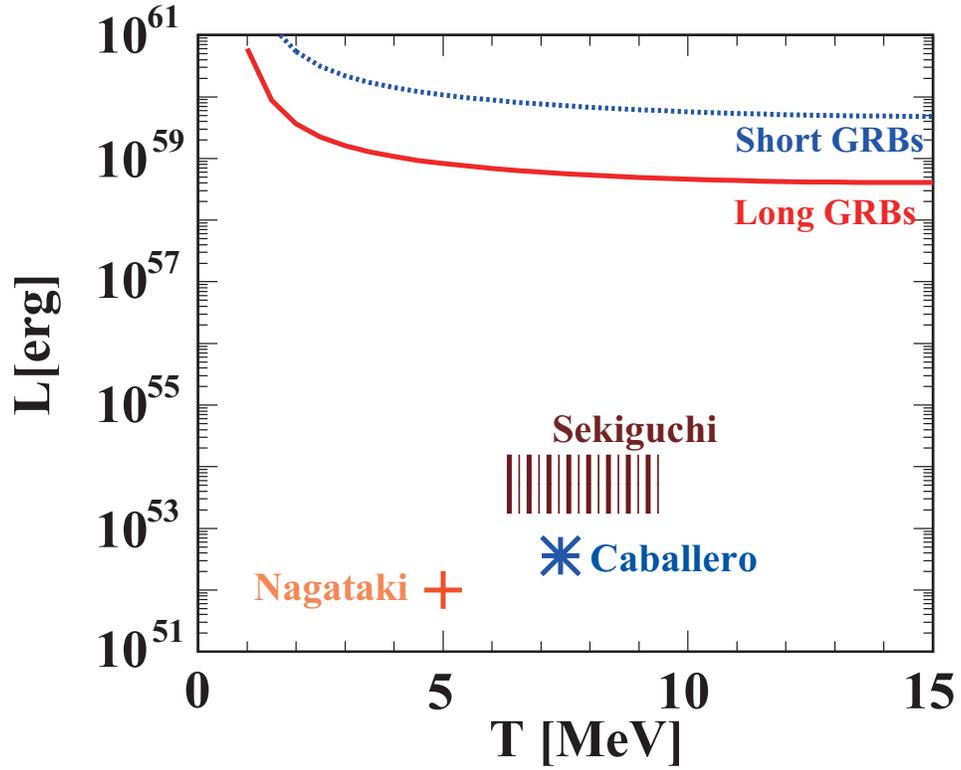}
\end{center}
\caption{The KamLAND constraint on the luminosity ($L$) and effective temperature ($T$) 
relation in the GRB accretion disk together with several 
theoretical calculations~\citep{nagataki2002,sekiguchi2011,Caballero2009}. }
\label{fig5}
\end{figure}

$N_{90}$ can be translated to constrain the $\overline{\nu}_e$'s  
luminosity ($L$) and effective temperature ($T$) in the accretion disk 
using the assumption that the $\overline{\nu}_e$ flux follows the Fermi-Dirac distribution described:
\begin{equation}
 \psi(E_{\overline{\nu}_e},T,L) = \frac{120}{7\pi^4}\frac{L}{T^4}\frac{E_{\overline{\nu}_e}^2}{\exp(E_{\overline{\nu}_e}/T)+1}. 
\end{equation} 
The expected total flux at the detector is in Period $k$, 
\begin{equation}
  \Psi^k(E_{\overline{\nu}_e},T,L)=\sum_{i \in k} \frac{1+z_i}{4\pi d^2_i}\psi((1+z_i)E_{\overline{\nu}_e},T,L), 
\end{equation}
where $z_i$ and $d_i$ are the redshift and luminosity distance of the $i$\,th GRB. 
The luminosity and effective temperature upper limits ($T_{\rm up}$, $L_{\rm up}$) 
are then connected to $N_{90}$: 
\begin{equation}
 N_{90} = \sum_{k} \int_{E^k_{\rm low}}^{100\,{\rm MeV}} I'_k(T_{\rm up},L_{\rm up},E^{\rm vis}_{\rm p}) dE^{\rm vis}_{\rm p},
\end{equation}
where $I'_k$ is the visible spectrum of the DC events: 
\begin{equation}
 I_k'(T_{\rm up},L_{\rm up},E^{\rm vis}_{\rm p})=\int_{E^k_{\rm low}}^{100\,{\rm MeV}} N_{\rm T} \epsilon_t \epsilon^k_s(E^{\rm vis}_{\rm p}) \sigma_{\rm IBD}(E_{\overline{\nu}_e})  \Psi^k( E_{\overline{\nu}_e},T_{\rm up},L_{\rm up})
R(E^{\rm exp}_{\rm p},E^{\rm vis}_{\rm p})dE^{\rm exp}_{\rm p}. 
\end{equation}
With the assumption of $E_{\overline{\nu}_e} =  E_{\rm p}  + \delta E$, 
the results obtained from KamLAND are shown in Figure \ref{fig5}. 
The upper limit spectra ($\sum I_k'(T_{\rm up},L_{\rm up},E^{\rm vis}_{\rm p})$) with $T_{\rm up}=5, 10, 15$\,MeV 
are shown in Figure~\ref{fig3}. 

The limits are six orders of magnitude higher than the supernovae $\overline{\nu}_e$ luminosity 
and several orders of magnitude higher than theoretical studies predict. 
\cite{nagataki2002} analytically show that a collapsar emits $\overline{\nu}_e$ 
with $L=10^{52}$\,erg and $T=5$\,MeV in a total accretion mass of 30\,M$_{\sun}$, 
a initial mass of 3\,M$_{\sun}$, and a mass accretion rate of 0.1\,M$_{\sun}$/s. 
\cite{Caballero2009} numerically predict $L=3.5\times 10^{52}$\,erg during 0.15\,sec with $T=7.5$\,MeV for 
black hole-neutron star mergers. Here, we assumed the averaged $\overline{\nu}_e$ energy corresponds to $3.15T$. 
Recently, Sekiguchi presented $\dot{L}=1.5$--$3 \times 10^{52}$\,erg/s during 2--3\,sec 
with an averaged $\overline{\nu}_e$ energy of 20--30\,MeV for a merger of binary neutron stars using 
state-of-the-art numerical simulations~\citep{sekiguchi2011}. 

Finally, we comment about the approximation, $E_{\overline{\nu}_e} = E^{\rm exp}_{\rm p} + \delta E$. 
Above 20\,MeV, this approximation is not suitable. 
In addition, the effect of the recoiling neutron ($\overline{E}_{\rm n}$) to $E_{\rm p}$ is no longer negligible. 
This effect adds a substantial energy bias, $\sim 10$\%, 
but the uncertainty of $L_{\rm up}$ is much smaller than 10\%.  
The amount of the error has no impact on our result and discussion.

\section{Summary}
We find no evidence for $\overline{\nu}_e$ associated with our sample of GRBs in KamLAND. 
We placed the lowest observational bound on the $\overline{\nu}_e$ fluence below 7\,MeV. 
The relation of $L$-$T$, which characterizes the GRB accretion disk, is constrained. 
The obtained upper limits are significantly  higher than several theoretical predictions~\citep{nagataki2002,sekiguchi2011}. 
However, our result is the first constraint 
that can be directly compared to theoretical studies.  


\acknowledgments
We are indebted to the observers of GRBs for providing us with data. 
KamLAND is supported by MEXT KAKENHI Grant Numbers 26104002, 26104007; 
the World Premier International Research Center Initiative (WPI Initiative), MEXT, Japan; 
and under the U.S. Department of Energy (DOE) grants no. DE-FG03-00ER41138, DE-AC02-05CH11231, and DE-FG02-01ER41166, 
as well as other DOE grants to individual institutions, 
and Stichting Fundamenteel Onderzoek der Materie (FOM) in the Netherlands. 
The Kamioka Mining and Smelting Company has provided service for activities in the mine.
We thank the support of NII for SINET4.



\bibliographystyle{apj}
\bibliography{GRB}

\clearpage

\end{document}